\documentclass[pra,aps,twocolumn,showpacs]{revtex4}
\usepackage{times,epsfig,amssymb,amsfonts,amsmath}%\usepackage{epsfig}
\newcommand{\ket}[1]{|#1\rangle}
\newcommand{\bra}[1]{\langle #1|}
\newcommand{\proj}[1]{\ket{#1}\bra{#1}}

\begin{document}

\title{Hierarchical monogamy relations for the squared entanglement of formation in multipartite systems}

\author{Yan-Kui Bai$^{1,2}$}
\email{ykbai@semi.ac.cn}
\author{Yuan-Fei Xu$^{1}$}
\author{Z. D. Wang$^2$}
\email{zwang@hku.hk}

\affiliation{$^1$ College of Physical Science and Information
Engineering and Hebei Advance Thin Films Laboratory, Hebei Normal
University, Shijiazhuang, Hebei 050024, China\\
$^2$ Department of Physics and Centre of Theoretical and Computational
Physics, The University of Hong Kong, Pokfulam Road, Hong Kong, China}

%%%%%%%%%%%%%%%%%%%%%%%%%%%%%%%%%%%%%%%%%%%%%%%%%%%%%%%%%%%%%%%%%%%%%%%%%%%%%%%%%%%%%%%%%%%%%%%%%
\begin{abstract}
We show exactly that the squared entanglement of formation (SEF) obeys a set of hierarchical monogamy
relations for an arbitrary $N$-qubit mixed state. Based on this set of monogamy relations, we are
able to construct the set of hierarchical multipartite entanglement indicators for $N$-qubit states,
which still work well even when the concurrence-based indicators lose efficacy.
As a by-product, an intriguing analytical relation between the entanglement of formation (EOF) and
squared concurrence (SC) for an arbitrary mixed state of $2\otimes d$ systems is derived, making
the concurrence calculable via the corresponding EOF. Furthermore, we analyze the
multipartite entanglement dynamics in composite cavity-reservoir systems with the present set of
hierarchical indicators. Moreover, for multilevel systems, it is illustrated that the SEF can be
monogamous even if the SC is polygamous.
\end{abstract}
%%%%%%%%%%%%%%%%%%%%%%%%%%%%%%%%%%%%%%%%%%%%%%%%%%%%%%%%%%%%%%%%%%%%%%%%%%%%%%%%%%%%%%%%%%%%%%%%%


\maketitle

\section{Introduction}

Entanglement monogamy is one of the most important properties for many-body quantum systems
\cite{horo09rmp, jens14jpa}, which means that quantum entanglement cannot be freely shared among many
parties and there is a trade-off among the amount of entanglement in different subsystems. For example,
in a three-qubit system $\rho_{ABC}$, when qubits $A$ and $B$ are maximally entangled, the third qubit
$C$ cannot be correlated with qubits $AB$ at all \cite{ben96pra1}. For entanglement quantified by
the squared concurrence (SC) \cite{woo97prl}, Coffman \emph{et al.} proved the first quantitative
monogamy relation for three-qubit states \cite{ckw00pra}, \emph{i.e.},
$C^2(\rho_{A|BC})-C^2(\rho_{AB})-C^2(\rho_{AC})\geq 0$, and the residual entanglement can be used
to characterize the genuine three-qubit entanglement \cite{dur00pra}. Furthermore, Osborne and
Verstraete proved the corresponding relation for $N$-qubit systems \cite{osborne06prl}
$C^2(\rho_{A_1|A_2\cdots A_N})-C^2(\rho_{A_1A_2})-\cdots-C^2(\rho_{A_1A_N})\geq0$, in which
$C^2(\rho_{A_1|A_2\cdots A_N})$ characterizes bipartite entanglement in the partition
$A_1|A_2\cdots A_N$ and $C^2(\rho_{A_1A_i})$ quantifies two-qubit entanglement with $i=2,3,\cdots, N$.
As is known, the entanglement monogamy property can be used to characterizing multipartite
entanglement structure \cite{ckw00pra,dur00pra,osborne06prl}, based on which some multipartite
entanglement measures and indicators are introduced and utilized to detect the existence of
multiqubit entanglement in dynamical procedures
\cite{byw07pra,baw08pra,eos09pra,csyu05pra,ouf08pra,byw09pra,ren10pra,cor13pra,agu14arxiv}.
Similar monogamy relations were also generalized to Gaussian systems \cite{adesso06njp,hir07prl},
squashed entanglement \cite{koashi04pra,chr04jmp}, entanglement negativity
\cite{fan07pra,kim09pra,hhe14arxiv}, entanglement Renyi entropy \cite{kim10jpa,cor10pra}, Lorentz
invariance \cite{jens14arxiv} and strong entanglement monogamy cases \cite{ads07prl,reg14prl,kim14pra}.

The entanglement of formation (EOF) is also a well-defined bipartite entanglement measure and has
operational meaning in entanglement preparation and data storage \cite{ben96pra1}. Unfortunately,
EOF itself does not satisfy the usual monogamy relation even for three-qubit pure states. Recently,
it was indicated that the squared entanglement of formation (SEF) obeys the monogamy relation in multiqubit
systems \cite{bai13pra,oli14pra,bxw14prl,zhu14pra}. In particular, we proved analytically that the SEF
is monogamous in an arbitrary $N$-qubit mixed state and obeys the relation \cite{bxw14prl}
\begin{equation}\label{1}
E_f^2(\rho_{A_1|A_2\cdots A_N})-E_f^2(\rho_{A_1A_2})-\cdots -E_f^2(\rho_{A_1A_N})\geq 0.
\end{equation}
In comparison with the $N$-qubit monogamy relation for SC \cite{osborne06prl},
the advantages of that for SEF shown in Eq. (1) are that (i) the residual entanglement of SEF
can indicate all multiqubit entangled states in the $N$-partite systems \cite{bxw14prl} and (ii) unlike
the concurrence $C(\rho_{A_1|A_2\cdots A_N})$, the multiqubit EOF $E_f(\rho_{A_1|A_2\cdots A_N})$
can be calculated via quantum discord \cite{oll01prl,ved01jpa,datta13ijmpb,fran13ijmpb,modi12rmp}
without resorting to the convex roof extension \cite{ben96pra2}.

Osborne and Verstraete  proved  that  When  an  $N$-qubit quantum state is divided into $k$ parties, Osborne and Verstraete proved that
the SC obeys a set of hierarchical monogamy relations~\cite{osborne06prl},
\begin{equation}\label{2}
C^2(\rho_{A_1|A_2\cdots A_N})\geq \sum_{i=2}^{k-1} C^2(\rho_{A_1A_i})+C^2(\rho_{A_1|A_k\cdots A_N}),
\end{equation}
which can be used to detect the multipartite entanglement in $k$-partite cases with
$k=\{3,4,\cdots, N\}$. However, calculating multiqubit concurrence is extremely hard due to
the convex roof extension \cite{ben96pra2}, which makes the quantitative characterization on this
set of monogamy relations very difficult. Since the $N$-partite monogamy relation of SC has an
intrinsic relation with that of SEF \cite{bxw14prl}, it is natural to ask whether or not the SEF in
$k$-partite systems satisfies similar hierarchical monogamy relations, considering that the
bipartite multiqubit EOF is calculable via effective methods for calculating quantum discord
\cite{luo08pra,lang10prl,gio10prl,ade10prl,ali10pra,huang13pra,cen11pra,che11pra,boli11pra,shi12pra}.
Moreover, is the amount of EOF related to that of SC in multiqubit systems (and, if so, how)?
On the other hand, whether the monogamy properties of SEF
and SC are equivalent in general multipartite systems still seems to be a fundamental open question.

In this paper, we show exactly that the SEF obeys a set of hierarchical $k$-partite monogamy relations
in an arbitrary $N$-qubit mixed state. Based on these monogamy relations, a set of hierarchical
multipartite entanglement indicators which can still work well
even when the concurrence-based indicators lose their efficacy is constructed correspondingly. As a byproduct, we also obtain
the analytical relation between EOF and SC in $2\otimes d$ systems. Furthermore, we analyze the
multipartite entanglement dynamics in cavity-reservoir systems with the presented hierarchical
indicators. Finally, we make a comparative study of the monogamy properties of SEF and SC, which are
inequivalent in multi-level systems.

\section{Hierarchical $K$-partite monogamy relations for SEF in $N$-qubit systems}
In a bipartite mixed state $\varrho_{AB}$, the EOF is defined as \cite{ben96pra2}
\begin{equation}\label{3}
E_f(\varrho_{AB})=\mbox{min}\sum_i p_i E_f(\ket{\psi^i}_{AB}),
\end{equation}
where $E_f(\ket{\psi^i}_{AB})=S(\rho_A^i)=-\mbox{Tr}\rho_A^i\mbox{log}_2\rho_A^i$ is the von
Neumann entropy and minimum running over all the pure
state decompositions. In particular, for two-qubit states, an analytical formula was given by
Wootters \cite{woo98prl}
\begin{equation}\label{4}
E_f(\rho_{AB})=h(\frac{1+\sqrt{1-C_{AB}^2}}{2}),
\end{equation}
where $h(x)=-x\mbox{log}_2x-(1-x)\mbox{log}_2(1-x)$ is the binary entropy and
$C_{AB}=\mbox{max}\{0,\sqrt{\lambda_1}-\sqrt{\lambda_2}-\sqrt{\lambda_3}-\sqrt{\lambda_4}\}$ is
the concurrence, with $\lambda_i$ being the decreasing eigenvalues of matrix
$\rho_{AB}(\sigma_y\otimes \sigma_y)\rho_{AB}^*(\sigma_y\otimes \sigma_y)$.

In this work, a key result is to show exactly a set of hierarchical $k$-partite monogamy relations
for SEF in arbitrary $N$-qubit mixed states $\rho_{A_1A_2\cdots A_N}$,
\begin{equation}\label{5}
E_f^2(\rho_{A_1|A_2\cdots A_N})\geq \sum_{i=2}^{k-1}E_f^2(\rho_{A_1A_i})
+E_f^2(\rho_{A_1|A_k\cdots A_N})
\end{equation}
for $k=\{3,4,\cdots N\}$, where the relation for $k=N$ is just the above-mentioned monogamy inequality
of Eq. (1). To show this set of monogamy relations of SEF, we first  prove the following
lemmas and theorems.

\emph{Lemma 1}. For two-qubit mixed states, the entanglement of formation $E_f(C^2)$ is monotonic
and concave as a function of squared concurrence $C^2$.

\emph{Proof}. The monotonically increasing property of $E_f(C^2)$ is satisfied if the first-order
derivative $dE_f/dx>0$, with $x=C^2$. According to Eq. (4), we have
\begin{equation}\label{6}
\frac{dE_{f}}{dx}=\frac{1}{\sqrt{1-x}\cdot \mbox{ln}16}\mbox{ln}(\frac{1+\sqrt{1-x}}{1-\sqrt{1-x}}).
\end{equation}
When $x\in (0,1)$, the first-order derivative is positive. Combining this fact with the observation that
$E_f(0)=0$ and $E_f(1)=1$ correspond respectively to its minimum and maximum, we can deduce that
$E_f$ is a monotonically increasing function of $x$. Moreover, the concave property of $E_f(x)$ holds
if the second-order derivative $d^2E_f/dx^2<0$. After some deduction, we have
\begin{eqnarray}\label{7}
\frac{d^{2}E_{f}}{dx^2}=g(x)\cdot \{-2\sqrt{1-x}+x\cdot
\mbox{ln}(\frac{1+\sqrt{1-x}}{1-\sqrt{1-x}})\}<0,
\end{eqnarray}
where $g(x)=1/[2\cdot(\mbox{ln}16)\cdot x(1-x)^{3/2}]$ is a nonnegative factor. Therefore, the
entanglement of formation $E_f$ is a concave function of $x$. The details for illustrating the
negativity of Eq. (7) are presented in Appendix A.

\emph{Lemma 2}. For two-qubit mixed states, the entanglement of formation $E_f(C)$ is monotonic
and convex as a function of concurrence $C$.

\emph{Proof}. The monotonically increasing property of $E_f(C)$ is satisfied if the first-order
derivative $dE_f/dC>0$. According to Eq. (4), we have
\begin{equation}\label{8}
\frac{dE_{f}}{dC}=\frac{C}{\sqrt{1-C^2}\cdot \mbox{ln}4}\mbox{ln}(\frac{1+\sqrt{1-C^2}}{1-\sqrt{1-C^2}}).
\end{equation}
When $C\in (0,1)$, the first-order derivative is positive. Combining this fact with an observation that
$E_f(0)=0$ and $E_f(1)=1$ correspond respectively to its minimum and maximum, we can deduce that
$E_f$ is a monotonically increasing function of $C$. Furthermore, the convex property of $E_f(C)$ holds
if the second-order derivative $d^2E_f/dC^2>0$. After some deduction, we have
\begin{eqnarray}\label{9}
\frac{d^{2}E_{f}}{dC^2}=u(C)\cdot \{-2\sqrt{1-C^2}+
\mbox{ln}(\frac{1+\sqrt{1-C^2}}{1-\sqrt{1-C^2}})\}>0,
\end{eqnarray}
where $u(C)=1/[(\mbox{ln}4)\cdot (1-C^2)^{3/2}]$ is a non-negative factor. Therefore, the
entanglement of formation $E_f$ is a convex function of $C$. The details for proving the
positivity of Eq. (9) are shown in Appendix B.

\emph{Theorem 1}. For a bipartite $2\otimes d$ mixed state $\rho_{AC}$, the entanglement of formation
obeys the following relation
\begin{equation}\label{10}
E_f(\rho_{AC})= E_f[C^2(\rho_{AC})],
\end{equation}
where the function on the right-hand side has the same expression as that of two-qubit EOF shown in
Eq. (4) with $C^2$ being the squared concurrence of $2\otimes d$ systems.

\emph{Proof}. According to Eq. (3), the EOF in $2\otimes d$ systems has the form
$E_f(\rho_{AC})=\mbox{min}\sum_i p_i E_f(\ket{\psi^i}_{AC})$. Under the optimal pure state
decomposition $\{p_i, \ket{\psi^i}_{AC}\}$, we have
\begin{eqnarray}\label{11}
E_f(\rho_{AC})&=&\sum_i p_i E_f(\ket{\psi^i}_{AC})\nonumber\\
&=&\sum_i p_i E_f[C^2(\ket{\psi^i}_{AC})]\nonumber\\
&\leq& \sum_j q_j E_f[C^2(\ket{\varphi^j}_{AC})]\nonumber\\
&\leq& E_f[\sum_j q_j C^2(\ket{\varphi^j}_{AC})]\nonumber\\
&=&E_f[C^2(\rho_{AC})],
\end{eqnarray}
where  we have used in the second equality the Wootters formula for pure states since the $2\otimes d$
pure state $\ket{\psi^i}_{AC}$ is equivalent to a two-qubit state under Schmidt decomposition
\cite{peres95book} and have taken the EOF $E_f(\ket{\psi^i}_{AC})$ as a function of the squared concurrence
$C^2(\ket{\psi^i}_{AC})$, and in the third inequality  the optimal decomposition
$\{q_j,\ket{\varphi^j}_{AC}\}$ for the concurrence $C^2(\rho_{AC})=\mbox{min}\sum_j q_j
C^2(\ket{\varphi^j}_{AC})$ \cite{osborne06prl}, which results in the average EOF being not less than
$E_f(\rho_{AC})$. The fourth inequality holds because of the concave property of $E_f(C^2)$ as proved
in Lemma 1, and the last equality is satisfied because $\{q_j,\ket{\varphi^j}_{AC}\}$ is the optimal
pure-state decomposition for $C^2(\rho_{AC})$. On the other hand, under the optimal pure-state
decomposition of $E_f(\rho_{AC})$, we also have
\begin{eqnarray}\label{12}
E_f(\rho_{AC})&=&\sum_i p_i E_f(\ket{\psi^i}_{AC})\nonumber\\
&=&\sum_i p_i E_f[C(\ket{\psi^i}_{AC})]\nonumber\\
&\geq& E_f[\sum_i p_i C(\ket{\psi^i}_{AC})]\nonumber\\
&\geq& E_f[\sum_k r_k C(\ket{\phi^k}_{AC})]\nonumber\\
&=&E_f[C(\rho_{AC})],
\end{eqnarray}
where, we have used in the second equality the Wootters formula for pure states and have taken the EOF
$E_f(\ket{\psi^i}_{AC})$ as a function of the concurrence $C(\ket{\psi^i}_{AC})$, in the third
inequality we have used the convex property of $E_f(C)$ (proved in Lemma 2) as a function of concurrence $C$, and
in the fourth inequality we have used the optimal pure-state decomposition $\{r_k, \ket{\phi^k}_{AC}\}$ for the
concurrence $C(\rho_{AC})$ and the monotonically increasing property of $E_f(C)$. Combining Eq. (11)
with Eq. (12), we can obtain
\begin{equation}\label{13}
E_f[C(\rho_{AC})]\leq E_f(\rho_{AC})\leq E_f[C^2(\rho_{AC})].
\end{equation}
Furthermore, according to the Wootters formula in Eq. (4), we have
\begin{equation}\label{14}
E_f[C(\rho_{AC})]=h(\frac{1+\sqrt{1-C_{AC}^2}}{2})=E_f[C^2(\rho_{AC})],
\end{equation}
where $E_f[C(\rho_{AC})]=E_f[C^2(\rho_{AC})]$ since they have the same expression, with $h(x)$ being the
binary entropy function. Therefore, the inequality signs in Eq. (13) become equality signs, and
then Theorem 1 is satisfied.

\emph{Theorem 2}. For a tripartite mixed state $\rho_{ABC}$ of $2\otimes 2\otimes 4$ systems, the
squared entanglement of formation obeys the monogamy relation
\begin{equation}\label{15}
E_f^2(\rho_{A|BC})-E_f^2(\rho_{AB})-E_f^2(\rho_{AC})\geq 0,
\end{equation}
where $\rho_{AB}$ and $\rho_{AC}$ are the reduced quantum states of $2\otimes 2$ and $2\otimes 4$
subsystems, respectively.

\emph{Proof}. We first analyze the pure state case. In a tripartite pure state $\ket{\psi_{ABC}}$
of the $2\otimes 2\otimes 4$ systems, we can derive
\begin{eqnarray}\label{16}
&&E_f^2(\ket{\psi_{A|BC}})-E_f^2(\rho_{AB})-E_f^2(\rho_{AC})\nonumber\\
&=& E_f^2[C^2(\ket{\psi_{A|BC}})]-E_f^2[C^2(\rho_{AB})]-E_f^2[C^2(\rho_{AC})]\nonumber\\
&\geq& E_f^2(C^2_{AB}+C^2_{AC})-E_f^2(C^2_{AB})-E_f^2(C^2_{AC})\nonumber\\
&\geq& 0,
\end{eqnarray}
where we have used the property $E_f(\rho_{AC})=E_f[C^2(\rho_{AC})]$ as proved in Theorem 1
and the property of $\ket{\psi_{ABC}}$ being equivalent to a two-qubit state under the partition
$A|BC$ in the first equality,  the monotonic property of $E_f^2(C^2)$ and the monogamy relation
$C^2_{A|BC}\geq C^2_{AB}+C^2_{AC}$~\cite{osborne06prl} in the second inequality,
and the convexity of function $E_f^2(C^2)$ \cite{bxw14prl} in the last
inequality. Thus, we have proven the
monogamy relation for pure state cases. Next, we prove it for mixed states.
The EOF in bipartite partition $A|BC$ is $E_f(\rho_{A|BC})=\mbox{min}\sum_ip_i E_f(\ket{\psi^i_{A|BC}})$
with the minimum running over all pure-state decompositions. Under the optimal decomposition
$\{p_i,\ket{\psi^i_{ABC}}\}$, we can get
\begin{eqnarray}\label{17}
&&E_f(\rho_{A|BC})=\sum_i p_i E_f(\ket{\psi^i_{A|BC}})=\sum_i E1_i\nonumber\\
&&E_f^\prime(\rho_{AJ})=\sum_i p_i E_f(\rho_{AJ}^i)=\sum_i Ej_i,
\end{eqnarray}
where $E_f^\prime(\rho_{AJ})$ (with $J=B,C$ and $j=2,3$) is the average EOF under the specific
decomposition. Consequently, we can derive
\begin{eqnarray}\label{18}
&&E_f^2(\rho_{A|BC})-E_f^{\prime 2}(\rho_{AB})-E_f^{\prime 2}(\rho_{AC})\nonumber\\
&=& (\sum_i E1_i)^2-(\sum_i E2_i)^2-(\sum_i E3_i)^2\nonumber\\
&=& \sum_i(E1_i^2-E2_i^2-E3_i^2)+\Delta \geq 0,
\end{eqnarray}
where, in the second equation, the first term is nonnegative due to the proved monogamy relation in
the pure state case, and the second term $\Delta=2\sum_i\sum_{k=i+1}(E1_iE1_k-\sum_jEj_iEj_k)$
is also nonnegative from a rigorous analysis given in Appendix C.  On the other hand, we have
\begin{eqnarray}\label{19}
E_f(\rho_{AB})\leq E_f^{\prime}(\rho_{AB}),
E_f(\rho_{AC})\leq E_f^{\prime}(\rho_{AC})
\end{eqnarray}
since $E_f^{\prime}$ is a specific average EOF which is not less than that under the optimal pure state
decomposition. Combining Eqs. (18) and (19), we obtain the monogamy relation for mixed states, which
completes the proof of this theorem.

\emph{Theorem 3}. For an arbitrary tripartite quantum state $\rho_{AB\mathbb{C}}$ of $2\otimes 2\otimes
2^{N-2}$ systems, the monogamy relation $E_f^2(\rho_{A|B\mathbb{C}})-E_f^2(\rho_{AB})
-E_f^2(\rho_{A\mathbb{C}})\geq 0$ is satisfied.

\emph{Proof}. In a tripartite pure state $\ket{\psi_{AB\mathbb{C}}}$ of $2\otimes 2\otimes
2^{N-2}$ systems, the party $\mathbb{C}$ is equivalent to a logic four-level subsystem according to
the Schmidt decomposition \cite{peres95book} in the partition $AB|\mathbb{C}$. Therefore, the pure state
monogamy relation for this theorem is automatically satisfied in terms of the result of Theorem 2.
Also, we can prove it for the mixed state case. For the mixed state $\rho_{AB\mathbb{C}}$, we have
$E_f(\rho_{A|B\mathbb{C}})=\mbox{min}\sum_ip_iE_f(\ket{\psi_{A|B\mathbb{C}}^i})$, with the minimum
running over all the pure state decompositions. Under the optimal decomposition $\{p_i,
\ket{\psi_{AB\mathbb{C}}^i}\}$, we can obtain
$E_f(\rho_{A|B\mathbb{C}})=\sum_i p_i E_f(\ket{\psi_{A|B\mathbb{C}}^i})=\sum_i E1_i$,
$E_f^\prime(\rho_{AB})=\sum_i p_i E_f(\rho_{AB}^i)=\sum_i E2_i$, and
$E_f^\prime(\rho_{A\mathbb{C}})=\sum_i p_i E_f(\rho_{A\mathbb{C}}^i)=\sum_i E3_i$, in which
$E_f^\prime(\rho_{AB})$ and $E_f^\prime(\rho_{A\mathbb{C}})$ are the average entanglement in the
specific decomposition. We thus have
\begin{eqnarray}\label{20}
&&E_f^2(\rho_{A|B\mathbb{C}})-E_f^{\prime 2}(\rho_{AB})-E_f^{\prime 2}(\rho_{A\mathbb{C}})\nonumber\\
&=&\sum_i p_i^2[E_f^2(\ket{\psi_{A|B\mathbb{C}}^i})-E_f^{2}(\rho_{AB}^i)
-E_f^{2}(\rho_{A\mathbb{C}}^i)]\nonumber\\
&&+2\sum_i\sum_{k=i+1}(E1_iE1_k-\sum_{j=2}^3 Ej_iEj_k),
\end{eqnarray}
where the first term is nonnegative since the monogamy relation is satisfied for the pure-state case,
and  the second term is nonnegative as well, as shown in Appendix C.
Moreover, we have $E_f^{\prime}(\rho_{AB})\geq E_f(\rho_{AB})$ and $E_f^{\prime}(\rho_{A\mathbb{C}})
\geq E_f(\rho_{A\mathbb{C}})$ because $E_f^\prime$ is the average EOF under a specific decomposition.
Therefore, we have
\begin{equation}\label{21}
E_f^2(\rho_{A|B\mathbb{C}})-E_f^2(\rho_{AB})-E_f^2(\rho_{A\mathbb{C}})\geq 0.
\end{equation}

At this stage, we prove the hierarchical $k$-partite monogamy relations of SEF in an $N$-qubit mixed
state $\rho_{A_1A_2\cdots A_N}$. According to Theorem 3, the three-partite monogamy relation in the
$N$-qubit system is satisfied, and we have
\begin{equation}\label{22}
E_f^2(\rho_{A_1|A_2\cdots A_N})\geq E_f^2(\rho_{A_1A_2})+E_f^2(\rho_{A_1|A_3\cdots A_N}).
\end{equation}
Applying theorem 3 to the subsystem $\rho_{A_1|A_3\cdots A_N}$ again, we can derive the four-partite
monogamy relation $E_f^2(\rho_{A_1|A_2\cdots A_N})\geq E_f^2(\rho_{A_1A_2})+E_f^2(\rho_{A_1A_3})
+E_f^2(\rho_{A_1|A_4\cdots A_N})$. By the successive application of the Theorem 3 , we can
obtain a set of hierarchical $k$-partite monogamy relations for SEF with $k\in\{3,4,\cdots, N\}$,
such that we complete the whole proof for the monogamy inequalities shown in Eq. (5).

\section{Hierarchical indicators for multipartite entanglement}

For an $N$-qubit pure state $\ket{\psi_N}$, we are able to construct a set of hierarchical multipartite
entanglement indicators based on the corresponding monogamy relations for the SEF
\begin{eqnarray}\label{23}
\tau_{SEF(k)}(\ket{\psi_N})&=&E_f^2(\ket{\psi}_{A_1|A_2\cdots A_N})-\sum_{i=2}^{k-1}
E_f^2(\rho_{A_1A_i})\nonumber\\
&&-E_f^2(\rho_{A_1|A_k\cdots A_N}),
\end{eqnarray}
which can be used to detect multipartite entanglement for the $k$-partite case of an $N$-qubit system
under the partition $A_1|A_2\cdots A_N$. Moreover, for $N$-qubit mixed states, we can construct two
types of multipartite entanglement indicators
\begin{eqnarray}\label{24}
\tau_{SEF(k)}^{(1)}(\rho_N)&=&\mbox{min}\sum_i p_i \tau_{SEF(k)}(\ket{\psi^i_{N}}),\nonumber\\
\tau_{SEF(k)}^{(2)}(\rho_N)&=&E_f^2(\rho_{A_1|A_2\cdots A_N})-\sum_{i=2}^{k-1}E_f^2(\rho_{A_1A_{i}})\nonumber\\
&&-E_f^2(\rho_{A_1|A_k\cdots A_N}),
\end{eqnarray}
where the first type is based on the convex roof extension with the minimum running over all the pure-
state decompositions $\{p_i,\ket{\psi_N^i}\}$, while the second type comes from the mixed-state monogamy
relations for SEF. When the party number $k=N$, the two indicators in Eq. (24) are just the
multiqubit entanglement indicators introduced in Ref. \cite{bxw14prl}, which can detect entangled
multiqubit states without the concurrence and $n$-tangles \cite{loh06prl,bai08pra2}. The detection
ability of the first type of indicator is stronger than that of the second one, but the computability
of the second type is better since the bipartite multiqubit EOF can be obtained via quantum discord.

\begin{figure}
\epsfig{figure=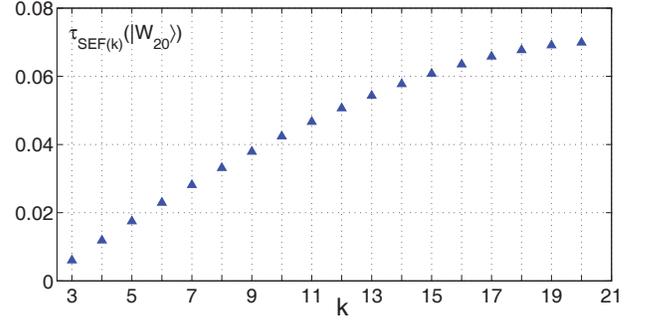,width=0.45\textwidth}
\caption{(Color online) The multipartite entanglement indicator $\tau_{SEF(k)}(\ket{W_{20}})$
as a function of party number $k$, where the nonzero values indicate the existence of multipartite
entanglement in the $k$-partite case of a $20$-qubit $W$ state with $k\in[3,20]$.}
\end{figure}

As an application, we first analyze the $N$-qubit $W$ state which has the form
\begin{equation}\label{25}
\ket{W_N}=\frac{1}{\sqrt{N}}(\ket{10\cdots 0}+\ket{01\cdots 0}+\cdots+\ket{00\cdots 1}).
\end{equation}
For this quantum state, the $k$-partite $N$-qubit monogamy relations of SC as shown in Eq. (2) are
saturated, and thus, the concurrence-based multipartite entanglement detection does not work. However,
we can use the newly introduced SEF-based indicator to represent the multipartite entanglement in
the $k$-partite case of $N$-qubit systems. After some deduction, we have
\begin{eqnarray}\label{26}
\tau_{SEF(k)}(\ket{W_N})&=&E_f^2(C_{A_1|A_2\cdots A_N}^2)-(k-2)E_f^2(C^2_{A_1A_2})\nonumber\\
&&-E_f^2(C^2_{A_1|A_k\cdots A_N}),
\end{eqnarray}
where we have used the symmetry of qubit permutations in the $W$ state and the squared concurrences are
$C^2_{A_1|A_2\cdots A_N}=4(N-1)/N^2$, $C^2_{A_1A_2}=4/N^2$, and $C^2_{A_1|A_k\cdots A_N}=4(N-k+1)/N^2$,
respectively. This set of $\tau_{SEF(k)}$ is positive since the SEF is a convex function of SC and
all the SCs in Eq. (26) are nonzero \cite{bxw14prl}. The nonzero $\tau_{SEF(k)}$ indicates the
existence of multipartite entanglement in the $k$-partite case of the $W$ state. In Fig£®1, we plot the
indicator as a function of party number $k$ in a $20$-qubit $W$ state, where the nonzero value detects
the multipartite entanglement in the $k$-partite case. Moreover, the hierarchy
relations are embodied by the monotonically increasing values of the indicator, with the minimum
$\tau_{SEF(3)}=0.00603$ and the maximum $\tau_{SEF(20)}=0.06989$, respectively.

\begin{figure}
\epsfig{figure=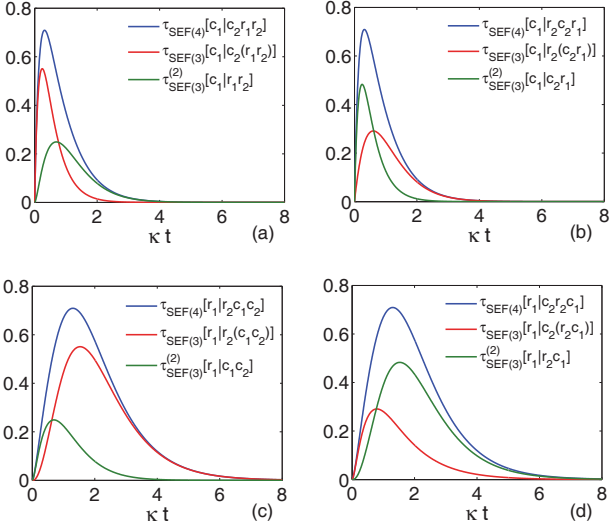,width=0.45\textwidth}
\caption{(Color online) The hierarchical multipartite entanglement indicators as functions of the time
evolution parameter $\kappa t$ in different partitions of composite cavity-reservoir systems:
(a) $c_1|c_2r_1r_2$, (b) $c_1|r_2c_2r_1$, (c) $r_1|r_2c_1c_2$, and (d) $r_1|c_2r_2c_1$.}
\end{figure}

Next, we make use of the hierarchical multipartite entanglement indicators to analyze a practical
dynamical procedure of two composite cavity-reservoir systems. The interaction of a single
cavity-reservoir system is described by the Hamiltonian
\cite{lop08prl,bel11ijqi,wen11epjd,bzww14epjd} $\hat{H}=\hbar
\omega\hat{a}^{\dagger}\hat{a}+\hbar\sum_{k=1}^{N}\omega_{k}
\hat{b}_k^{\dagger}\hat{b}_k+\hbar\sum_{k=1}^{N}g_{k}(\hat{a}
\hat{b}_{k}^{\dagger}+\hat{b}_{k}\hat{a}^{\dagger})$. When the initial state is
$\ket{\Phi_0}=(\alpha\ket{00}+\beta\ket{11})_{c_1c_2} \ket{00}_{r_1r_2}$, with the dissipative
reservoirs being in the vacuum state, the output state is equivalent to a four-qubit state and has
the form \cite{lop08prl}
\begin{equation}\label{27}
  \ket{\Phi_t}=\alpha\ket{0000}_{c_1r_1c_2r_2}+\beta\ket{\phi_t}_{c_1r_1}\ket{\phi_t}_{c_2r_2},
\end{equation}
where $\ket{\phi_t}=\xi(t)\ket{10}+\chi(t)\ket{01}$, with the amplitudes being $\xi(t)=\mbox{exp}
(-\kappa t/2)$ and $\chi(t)=[1-\mbox{exp}(-\kappa t)]^{1/2}$. Under the partition $c_1|c_2r_1r_2$,
the hierarchical multipartite entanglement indicators obey the following relation:
\begin{eqnarray}\label{28}
\tau_{SEF(4)}(\ket{\Phi_t}_{c_1|c_2r_1r_2})&=&\tau_{SEF(3)}(\ket{\Phi_t}_{c_1|c_2(r_1r_2)})\nonumber\\
&&+\tau_{SEF(3)}^{(2)}(\rho_{c_1|r_1r_2}),
\end{eqnarray}
where $\tau_{SEF(3)}$ and $\tau_{SEF(4)}$ are the multipartite entanglement indicators for the
three-partite and four-partite cases of the output state, and $\tau_{SEF(3)}^{(2)}$ detects the tripartite
entanglement in the three-qubit mixed state. In Fig. 2(a), we plot this set of hierarchical indicators as
functions of the time evolution $\kappa t$ (the initial-state parameter is chosen to be
$\alpha=1/\sqrt{3}$), where the nonzero values indicate the existence of multipartite entanglement.
Similarly, for the partitions $c_1|r_2c_2r_1$, $r_1|r_2c_1c_2$, and $r_1|c_2r_2c_1$, we can
also utilize the corresponding hierarchical SEF-based indicators to detect the multipartite entanglement
in pure and mixed states of the composite cavity-reservoir systems,  which are plotted in Fig. 2(b)-2(d).
Here, it should be pointed out that the advantage of SEF-based indicators is that the multiqubit
EOF of mixed states can be obtained via an effective method for calculating quantum discord
\cite{che11pra} (the details of the calculation are presented in Appendix D). However, for the
concurrence-based indicators, their calculation is very difficult because the bipartite multiqubit
concurrence in mixed states needs to resort to the convex-roof extension.

The hierarchy property of multipartite entanglement indicators $\tau_{SEF(k)}$ lies in not only the
values of different indicators but also the enhanced detection ability along with the party number $k$.
In Ref. \cite{bxw14prl}, it was proved that the nonzero indicator value of $N$-partite $N$-qubit
quantum states is a necessary and sufficient condition for the existence of multiqubit entanglement.
However, when one of the parties contains two or more qubits, the nonzero value of the indicator is only
a sufficient condition for multipartite entanglement detection, which is because there exist
multipartite entangled states with zero indicator values. As an example, we consider the tripartite
case of a four-qubit quantum state in $2\otimes 2\otimes 4$ systems,
\begin{equation}\label{29}
\ket{\psi}_{A_1A_2(A_3A_4)}=\frac{1}{2}(\ket{00\bar{0}}+\ket{10\bar{1}}+\ket{01\bar{2}}-\ket{11\bar{3}}),
\end{equation}
where the third party has two qubits $A_3$ and $A_4$ with the bases $\ket{\bar{0}}=\ket{00}$,
$\ket{\bar{1}}=\ket{01}$, $\ket{\bar{2}}=\ket{10}$, and $\ket{\bar{3}}=\ket{11}$, respectively. This
quantum state is multipartite entangled since the EOF is nonzero in any bipartite partition of the
tripartite system $A_1A_2(A_3A_4)$. But the tripartite entanglement indicator is
$\tau_{SEF(3)}(\ket{\psi}_{A_1|A_2(A_3A_4)})=0$, which fails to detect the multipartite entanglement
 [here, the EOF of reduced quantum state $\rho_{A_1(A_3A_4)}$ is $E_f(A_1|A_3A_4)=1$, since it is the
maximal entangled mixed state in $2\otimes 4$ systems \cite{cav05pra,zli12qic}]. On the other hand,
when the quantum state in Eq. (29) is taken to be a four-party case of a $2\otimes 2\otimes 2\otimes 2$
system, it has the form
\begin{equation}\label{30}
\ket{\psi}_{A_1A_2A_3A_4}=\frac{1}{2}(\ket{0000}+\ket{1001}+\ket{0110}-\ket{1111}).
\end{equation}
In this case, we can derive $\tau_{SEF(4)}(\ket{\psi})=1$, which indicates the existence of genuine
multipartite entanglement. In fact, this quantum state is just the four-qubit cluster state which is
genuinely four-partite entangled \cite{bri01prl,rau01prl}.

\section{Discussion}

Until now, the quantitative relation between the EOF and SC in general bipartite systems has been an
open problem. However, Theorem 1 in this paper provides an analytical expression for the relation of
EOF and SC in $2\otimes d$ systems, leading to the mixed-state concurrence in $2\otimes d$ systems
being available as long as we get the corresponding EOF. This is an important step forward in
calculating the mixed-state concurrence since the mixed-state EOF beyond two-qubit cases can be derived
via effective methods for calculating quantum discord
\cite{luo08pra,lang10prl,gio10prl,ade10prl,ali10pra,huang13pra,cen11pra,che11pra,boli11pra,shi12pra}.
For example, in the dynamical evolution of multipartite cavity-reservoir systems analyzed in Sec. III,
the calculation of SC $C^2_{c_1|r_1r_2}$ is extremely difficult according to the convex-roof extension.
But we can deduce this concurrence via the corresponding EOF. Based on the Koashi-Winter formula
\cite{koashi04pra} and the quantum discord in subsystem $c_1c_2$, we have the EOF (see details in
Appendix D)
\begin{equation}\label{31}
E_f(\rho_{c_1|r_1r_2})=-\eta_1\mbox{log}_2\eta_1-(1-\eta_1)\mbox{log}_2(1-\eta_1),
\end{equation}
where the parameter is $\eta_1=[1-(1-4\beta^2\xi^2\chi^2)^{1/2}]/2$. According to Theorem 1 in Sec. II,
we have the relation
\begin{equation}\label{32}
E_f(\rho_{c_1|r_1r_2})=h\left(\frac{1+\sqrt{1-C^2_{c_1|r_1r_2}}}{2}\right),
\end{equation}
where $h(x)=-x\mbox{log}_2x-(1-x)\mbox{log}_2(1-x)$ is the binary entropy. Combining Eqs. (31)
and (32), we can derive the SC
\begin{equation}\label{33}
C^2_{c_1|r_1r_2}=4\beta^2\xi^2\chi^2,
\end{equation}
where $\beta$ is the initial amplitude and the time evolution parameters are $\xi=\mbox{exp}(-\kappa t/2)$
and $\chi=[1-\mbox{exp}(-\kappa t)]^{1/2}$, respectively. In Fig. 3(a), we plot the concurrence
$C^2_{c_1|r_1r_2}$ as a function of time evolution $\kappa t$ and the initial amplitude $\alpha$, which
characterizes the evolution of SC in the dynamical procedure. Furthermore, in Fig. 3(b), the
entanglement distribution $C^2(\rho_{c_1|r_1r_2})-C^2(\rho_{c_1r_1})-C^2(\rho_{c_1r_2})$ is plotted,
which verifies the monogamy property of SC in the three-qubit mixed state.

\begin{figure}
\epsfig{figure=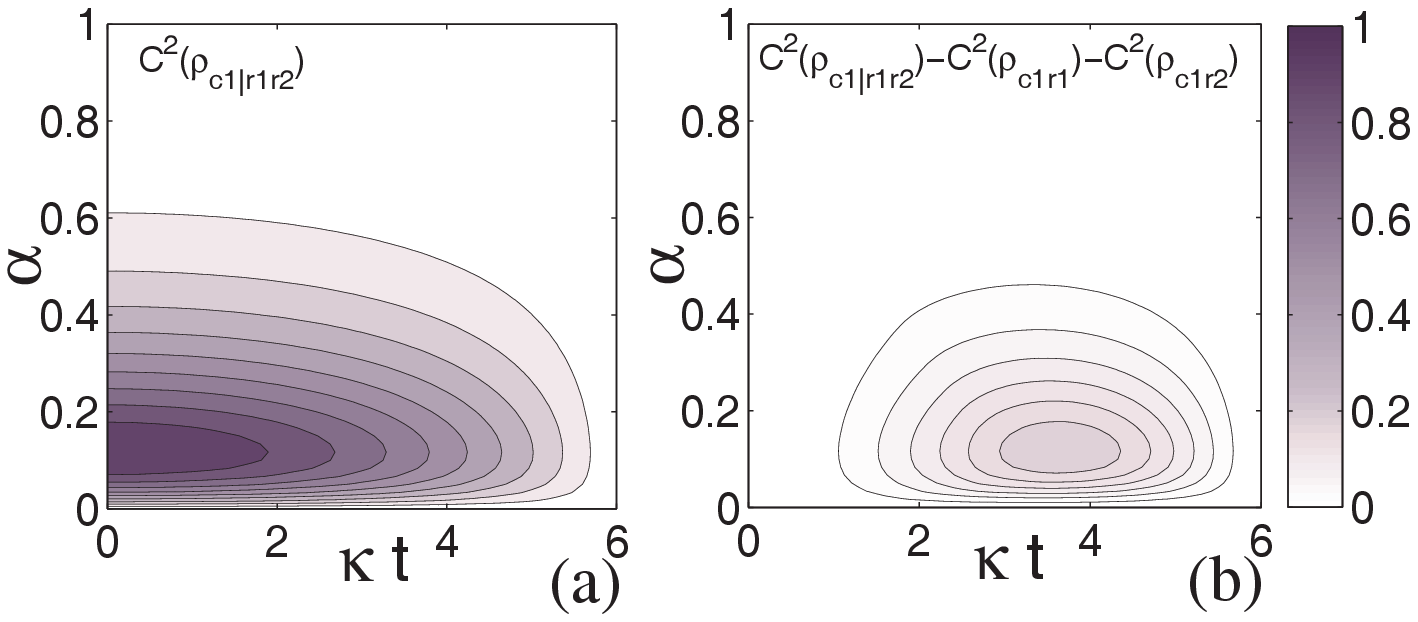,width=0.45\textwidth}
\caption{(Color online) Entanglement dynamics of squared concurrences as a functions of time evolution
$\kappa t$ and the initial amplitude $\alpha$:  (a) squared concurrence $C^2(\rho_{c_1|r_1r_2})$ and
(b) entanglement monogamy $C^2(\rho_{c_1|r_1r_2})-C^2(\rho_{c_1r_1})-C^2(\rho_{c_1r_2})$.}
\end{figure}

It was proven in Ref.~\cite{bxw14prl} that the SEF is monogamous in the $N$-partite case, as shown in
Eq. (1) where each party contains one qubit. In this paper, we further prove that the SEF monogamy is
satisfied even when the last party contains two or more qubits, and thus, we obtain a set of hierarchical
$k$-partite monogamy relations,
\begin{eqnarray}\label{34}
&&E_f^2(\rho_{A_1|A_2\cdots A_N})\nonumber\\
&\geq& E_f^2(\rho_{A_1A_2})+E_f^2(\rho_{A_1|A_3\cdots A_N})\nonumber\\
&\geq& E_f^2(\rho_{A_1A_2})+E_f^2(\rho_{A_1A_3})+E_f^2(\rho_{A_1|A_4\cdots A_N})\nonumber\\
&\vdots& ~~~~~~~~~~~~~~~~~~~~~\vdots~~~~~~~~~~~~~~~~~~~~~~\vdots~~~~~~~~~~~~~~~~~~~~~~\vdots\nonumber\\
&\geq& E_f^2(\rho_{A_1A_2})+\cdots +E_f^2(\rho_{A_1A_{N-2}})+E_f^2(\rho_{A_1|A_{N-1}A_N})\nonumber\\
&\geq& E_f^2(\rho_{A_1A_2})+\cdots+E_f^2(\rho_{A_1A_{N-1}})+E_f^2(\rho_{A_1A_N}),
\end{eqnarray}
where the  specific monogamy relation for $k=N$ reproduces the important result recently revealed in
Ref.~\cite{bxw14prl}. Note that the monogamy score is increasing along with the party number $k$
due to the hierarchy property of the inequalities, and on the basis of the hierarchy property we correspondingly 
present a set of multipartite entanglement indicators. For a general $N$-qubit mixed state $\rho_{N}$, the detection
ability of the multipartite entanglement indicator $\tau_{SEF(k)}$ is also enhanced along with the
increasing of party number $k$. When $k<N$, the nonzero value of the indicator is only a sufficient
condition for the existence of multipartite entanglement, but when $k=N$, the nonzero value is both
necessary and sufficient.

As shown in Eqs. (2) and (5), both SC and SEF satisfy the monogamy relation in multipartite $2\otimes
2\otimes 2 \cdots \otimes 2\otimes d$ systems. However, it is still an open question
whether or not the monogamy properties of SC and SEF are equivalent in multipartite systems of
\emph{arbitrary} dimension. Furthermore, does the SEF possess a \emph{better} monogamy property
than the SC?

First, we analyze multipartite $2\otimes d_2\otimes d_3 \cdots \otimes d_{N-1}\otimes d_N$ systems,
where only the first party is a two-level subsystem and the other parties are multilevel subsystems.
In this case, we have the following theorem, and the proof can be seen in Appendix E.

\emph{Theorem 4}. For multipartite $2\otimes d_2\otimes d_3 \cdots \otimes d_{N-1}\otimes d_N$ systems,
the monogamy property of squared entanglement of formation is superior to that of squared concurrence.

Here, it should be noted that the general monogamy property of SC is still an open problem for
$2\otimes d_2\otimes d_3 \cdots \otimes d_{N-1}\otimes d_N$ systems. However, according to the
Theorem 4, we know that the SEF in the multipartite systems should be monogamous whenever the SC
possesses this property, and furthermore, the SEF may still be monogamous even if the SC were
polygamous. As an example, we consider a four-partite mixed state
$\rho_{A_1\tilde{A}_2\tilde{A}_3\tilde{A}_4}$ of $2\otimes d_2\otimes d_3 \otimes d_4$ systems.
Suppose that the bipartite squared concurrences are $C^2_{A_1|\tilde{A}_2\tilde{A}_3\tilde{A}_4}=0.7$,
and $C^2_{A_1\tilde{A}_2}=C^2_{A_1\tilde{A}_3}=C^2_{A_1\tilde{A}_4}=0.3$.  Then we find that 
the SC is polygamous in this quantum state, \emph{i.e}.,
\begin{equation}\label{35}
C^2_{A_1|\tilde{A}_2\tilde{A}_3\tilde{A}_4}-\sum_{i=2}^4 C^2_{A_1\tilde{A}_i}=-0.2.
\end{equation}
However, if we use the SEF to characterize the entanglement distribution in this quantum state,
we can derive
\begin{eqnarray}\label{36}
&&E_f^2(\rho_{A_1|\tilde{A}_2\tilde{A}_3\tilde{A}_4})-\sum_{i=2}^4 E_f^2(\rho_{A_1\tilde{A}_i})\nonumber\\
&=&E_f^2(C^2_{A_1|\tilde{A}_2\tilde{A}_3\tilde{A}_4})-\sum_{i=2}^4 E_f^2(C^2_{A_1\tilde{A}_i})\nonumber\\
&=&0.594779-3\times 0.166494\nonumber\\
&=&0.0952982,
\end{eqnarray}
which is monogamous.

\begin{figure}
\epsfig{figure=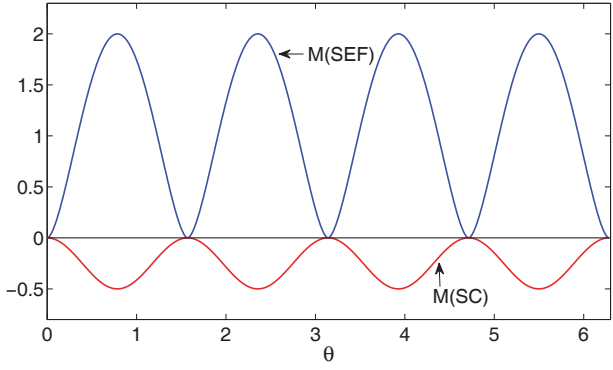,width=0.45\textwidth}
\caption{(Color online) The different distribution properties of
$M(SEF)=E_f^2(A|BC)-E_f^2(AB)-E_f^2(AC)$ and  $M(SC)=C^2(A|BC)-C^2(AB)-C^2(AC)$ as functions
of parameter $\theta$ in $4\otimes 2\otimes 2$ systems, where the SEF is monogamous and
the SC is polygamous.}
\end{figure}

Next, we investigate the monogamy properties of SEF and SC for multipartite quantum systems where
the first party is a multilevel subsystem. In Ref. \cite{ou07pra}, Ou indicated that the SC is
not monogamous for multipartite higher-dimensional systems, with a counterexample of $3\otimes
3\otimes 3$ systems being given by~\cite{ou07pra}
\begin{eqnarray}\label{37}
\ket{\Psi}_{ABC}&=&\frac{1}{\sqrt{6}}(\ket{123}-\ket{132}+\ket{231}-\ket{213}\nonumber\\
&&+\ket{312}-\ket{321}),
\end{eqnarray}
in which the monogamy score is $C^2_{A|BC}-C^2_{AB}-C^2_{AC}=-2/3$. However, when we use SEF to
characterize the entanglement distribution in this quantum state, we find that the corresponding
monogamy score is
\begin{eqnarray}\label{38}
&&E_f^2(\ket{\Psi}_{A|BC})-E_f^2(\rho_{AB})-E_f^2(\rho_{AC})\nonumber\\
&&=(\mbox{log}_23)^2-1-1\nonumber\\
&&\simeq 0.51211,
\end{eqnarray}
which is monogamous and indicates genuine tripartite entanglement. Therefore, the monogamy
properties of SEF and SC are inequivalent in multilevel systems.
Moreover, even when only the first party is multilevel, the monogamy properties of SEF and SC are
still different. As an example, we analyze a tripartite pure state of $4\otimes 2\otimes 2$ systems
\begin{equation}\label{39}
\ket{\Phi}_{ABC}=\frac{1}{\sqrt{2}}(\alpha\ket{000}+\beta\ket{110}+\alpha\ket{201}+\beta\ket{311}),
\end{equation}
where the parameters are $\alpha=\mbox{cos}\theta$ and $\beta=\mbox{sin}\theta$. 
In Fig. 4, we plot the monogamy properties of SEF and SC as functions of parameter $\theta$, and
it can be seen that the SEF is monogamous whereas the SC is polygamous (the details of the analysis of the
entanglement distribution property are given in Appendix F).

It is worth pointing out that a profound understanding of the monogamy property of SEF for a
general multipartite system is still lacking. From the above analysis on multi-level systems,
we may make the following two conjectures.

\emph{Conjecture 1}. For multipartite $2\otimes d_2\otimes d_3 \cdots \otimes d_{N-1}\otimes d_N$
systems, the squared entanglement of formation may be monogamous.

\emph{Conjecture 2}. For multipartite arbitrary $d$-dimensional quantum systems, the squared
entanglement of formation may be monogamous.

The proofs of these two conjectures are highly challenging and may demand some exotic tools for
characterizing the EOF in bipartite higher-dimensional systems, which is currently being explored
in the quantum information community.

\section{Conclusions}

To conclude,we have proven exactly that when an $N$-qubit quantum system is divided into $k$ parties, 
SEF obeys a set of hierarchical $k$-partite monogamy relations, as shown in Eq. (5), which is an
important generalization of the former $N$-partite $N$-qubit result \cite{bxw14prl}. In comparison
with the similar hierarchical monogamy properties of SC \cite{osborne06prl}, the merits of the SEF case
lie in its computability via quantum discord and its capability of multipartite entanglement detection.
Based on this set of monogamy relations for SEF, we are able to construct multipartite entanglement
indicators for various $k$-partite cases, which have a hierarchy structure and are still workable  even
when concurrence-based indicators lose their efficacy. In the evolution of four-partite
cavity-reservoir systems, the introduced indicators are utilized to analyze the dynamics of
multipartite entanglement, where a quantitative hierarchical relation between tripartite and
four-partite entanglement indicators is given in Eq. (28). Moreover, the hierarchy property of
multipartite entanglement indicators also lies in the improved detection ability along with the
increasing of party number $k$.

As an important by-product, we have also derived the analytical relation between EOF and SC in an
arbitrary mixed state of $2\otimes d$ systems (Theorem 1). This leads to the $2\otimes d$ concurrence being
computable without resorting to the convex-roof extension \cite{ben96pra2} since the EOF is
available via effective methods for calculating quantum discord \cite{luo08pra,lang10prl,gio10prl,
ade10prl,ali10pra,huang13pra,cen11pra,che11pra,boli11pra,shi12pra}. Therefore, beyond two-qubit cases,
the quantitative characterization of the monogamy relation of SC is possible. As an example, we
have calculated the entanglement distribution of SC in cavity-reservoir systems, which is plotted
in Fig. 3.

Finally, we have made a comparative study of the monogamy properties of SEF and SC in multilevel
systems. For multipartite $2\otimes d_2\otimes d_3\cdots d_{N-1}\otimes d_N$ systems, we have
proven that the monogamy property of SEF is superior to that of SC. When the first subsystem is
not a qubit, the concrete examples illustrate that the SEF can be monogamous even if the SC is
polygamous. However, in a general multipartite system, the monogamy property of SEF is still an
open problem, and proofs for the two conjectures are still needed.

\section*{Acknowledgments}

This work was supported by the RGC of Hong Kong under Grant Nos. HKU7058/11P and HKU7045/13P,
and the NFRPC (No. 2011CB922104). Y.-K.B. and Y.-F.X. were also supported by NSF-China
(Grant No. 10905016), Hebei NSF (Grant No. A2012205062), and the fund of Hebei Normal University.

\appendix

\section{Proof for the negativity of the second-order derivative in equation (7)}

In Eq. (7) of the main text, the second-order derivative has the form
\begin{eqnarray}\label{A1}
\frac{d^{2}E_{f}}{dx^2}=g(x)\cdot \{-2\sqrt{1-x}+x\cdot
\mbox{ln}(\frac{1+\sqrt{1-x}}{1-\sqrt{1-x}})\},
\end{eqnarray}
where $x=C^2$ and the factor $g(x)=1/[2\cdot(\mbox{ln}16)\cdot x(1-x)^{3/2}]$. Now we prove
that the derivative is negative.

In the region $x\in (0,1)$, the factor $g(x)$ is positive, and then the negativity of the derivative
is equivalent to $M(x)<0$, with
\begin{equation}\label{A2}
M(x)=-2\sqrt{1-x}+x\cdot \mbox{ln}(\frac{1+\sqrt{1-x}}{1-\sqrt{1-x}}).
\end{equation}
In order to determine the sign of $M(x)$, we first analyze the monotonic property of this function.
After some deduction, we find that the first-order derivative of $M(x)$ is
\begin{equation}\label{A3}
 \frac{dM(x)}{dx}=\mbox{ln}(\frac{1+\sqrt{1-x}}{1-\sqrt{1-x}}),
\end{equation}
which is positive since the term in the logarithm is larger than 1. Therefore, the function $M(x)$ is
monotonically increasing in the region $x\in (0,1)$. Next we analyze the values of $M(x)$ at two
end points. When $x=0$, we can get
\begin{eqnarray}\label{A4}
   \lim_{x \to +0}M(x)&=&\lim_{x \to
   +0}\{-2\sqrt{1-x}+x\cdot\mbox{ln}(\frac{1+\sqrt{1-x}}{1-\sqrt{1-x}})\}\nonumber\\
   &=&\lim_{x \to +0}-2\sqrt{1-x}+\lim_{x \to
   +0}\frac{\mbox{ln}(\frac{1+\sqrt{1-x}}{1-\sqrt{1-x}})}{\frac{1}{x}}\nonumber\\
   &=&-2+\lim_{x \to +0}\frac{x}{\sqrt{1-x}}\nonumber\\
   &=&-2,
\end{eqnarray}
where  we have used L'Hospital's rule in the third equation. When $x=1$, it is easy to obtain
$M(1)=0$. Combining the two end-point values with the monotonic property of $M(x)$, we have
$M(x)<0$ in the region $x\in(0,1)$, and thus, the second-order derivative $d^{2}E_{f}/dx^2<0$ in the
same region.

Furthermore, we analyze the second-order derivative at the end points. When $x=0$, we get
\begin{eqnarray}\label{A5}
  \lim_{x \to +0}\frac{d^{2}E_{f}}{dx^2}&=&\lim_{x \to +0}g(x)\cdot M(x)\nonumber\\
  &=&\lim_{x \to +0}g(x)\cdot\lim_{x \to +0}M(x)\nonumber\\
  &=&\infty\cdot(-2)\nonumber\\
  &=&-\infty,
\end{eqnarray}
where the result of Eq. (A4) has been used in the third equality. On the other hand, when $x=1$,
we can derive
\begin{eqnarray}\label{A6}
   \lim_{x \to 1}\frac{d^{2}E_{f}}{dx^2}&=&\lim_{x \to 1}\frac{M(x)}{\frac{1}{g(x)}}\nonumber\\
   &=&\lim_{x \to
   1}\frac{\mbox{ln}[(1+\sqrt{1-x})/(1-\sqrt{1-x})]}{(\mbox{ln}16)\cdot(2-5x)\sqrt{1-x}}\nonumber\\
   &=&\lim_{x \to 1}\frac{-1/(x\sqrt{1-x})}{[3\cdot(\mbox{ln}16)\cdot(5x-4)]/[2\sqrt{1-x}]}\nonumber\\
   &=&\frac{-2}{3\cdot\mbox{ln}16}\nonumber\\
   &\approx&-0.24.
\end{eqnarray}
Thus, we prove that the second-order derivative $d^{2}E_{f}/dx^2$ is negative in the whole region
$x\in[0,1]$ and complete the proof of Eq. (7) in the main text. In Fig. 5, we plot the
second-order derivative as a function of $x$, which illustrates our analytical result.

\begin{figure}
\epsfig{figure=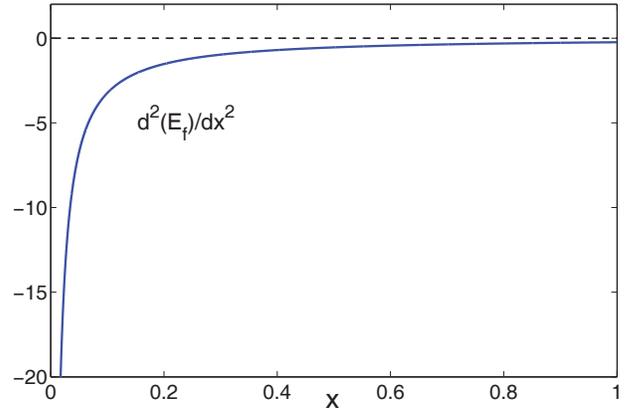,width=0.45\textwidth}
\caption{(Color online) The second-order derivative $d^{2}E_{f}/dx^2$ is plotted as a function of $x$,
with $x=C^2$, which is negative, and as a result, the EOF is a concave function of SC.}
\end{figure}

\section{Proof for the positivity of the second-order derivative in equation (9)}

In Eq. (9) of the main text, the second-order derivative has the form
\begin{eqnarray}\label{B1}
\frac{d^{2}E_{f}}{dC^2}=u(C)\cdot \{-2\sqrt{1-C^2}+
\mbox{ln}(\frac{1+\sqrt{1-C^2}}{1-\sqrt{1-C^2}})\},
\end{eqnarray}
where the factor is $u(C)=1/[(\mbox{ln}4)\cdot (1-C^2)^{3/2}]$. Now we prove that the derivative is
positive.

In the region $C\in (0,1)$, the factor $u(C)$ is positive and then the positivity of the derivative
is equivalent to $Q(C)>0$, with
\begin{equation}\label{B2}
Q(C)=-2\sqrt{1-C^2}+\mbox{ln}(\frac{1+\sqrt{1-C^2}}{1-\sqrt{1-C^2}}).
\end{equation}
In order to determine the sign of $Q(C)$, we first analyze the monotonic property of this function.
After some deduction, we find the first-order derivative of $Q(C)$ is
\begin{equation}\label{B3}
\frac{dQ(C)}{dC}=\frac{-2\sqrt{1-C^2}}{C},
\end{equation}
which is negative since the concurrence $C$ ranges in $(0,1)$. Therefore, the function $Q(C)$ is
monotonically decreasing in the region $C\in(0,1)$. Next, we investigate the values of $Q(C)$ at
two end points, which can be written as
\begin{eqnarray}\label{B4}
   \lim_{C \to +0}Q(C)&=&+\infty\nonumber\\
   \lim_{C \to 1}Q(C)&=&0.
\end{eqnarray}
Combining Eq. (\ref{B4}) with the monotonic property of $Q(C)$, we find that the function $Q(C)$
is positive in the region $C\in (0,1)$, and thus the second-order derivative $d^2E_f/dC^2>0$ in the
same region.

\begin{figure}
\epsfig{figure=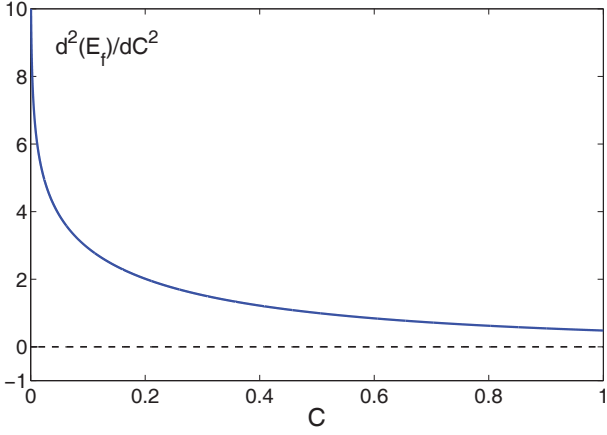,width=0.45\textwidth}
\caption{(Color online) The second-order derivative $d^{2}E_{f}/dC^2$ is plotted as a function of $C$, which is positive, and as a result, the EOF is a convex function of concurrence $C$.}
\end{figure}

Furthermore, we analyze the second-order derivative at the endpoints. When $C=0$, we have
\begin{eqnarray}\label{B5}
\lim_{C \to +0}\frac{d^2E_f}{dC^2}=\lim_{C \to +0}u(C)\cdot Q(C)=+\infty.
\end{eqnarray}
On the other hand, when $C=1$, we can derive
\begin{eqnarray}\label{B6}
\lim_{C \to 1}\frac{d^2E_f}{dC^2}&=&\lim_{C \to 1}u(C)\cdot Q(C)\nonumber\\
&=&\lim_{C \to 1}\frac{-2\sqrt{1-C^2}+\mbox{ln}(\frac{1+\sqrt{1-C^2}}{1-\sqrt{1-C^2}})}
{(1-C^2)^{3/2}\cdot \mbox{ln}4}\nonumber\\
&=&\lim_{C \to 1}\frac{1}{\mbox{ln}4}\cdot \frac{\frac{-2\sqrt{1-C^2}}{C}}{-3C\cdot\sqrt{1-C^2}}\nonumber\\
&=&\frac{2}{3\cdot\mbox{ln}4}\nonumber\\
&\approx &0.48.
\end{eqnarray}
Thus, we prove that the second-order derivative $d^2E_f/dC^2>0$ in the whole region $C\in[0,1]$ and
then complete the proof of Eq. (9) in the main text. In Fig.6, we plot the derivative as a function of
concurrence $C$, which verifies our analytical result.

\section{Nonnegative second terms in equation (18) and (20)}

We first analyze the term in Eq. (18) of the main text, which has the form
\begin{equation}\label{C1}
\Delta=2\sum_i\sum_{k=i+1}(E1_iE1_k-\sum_{j=2}^{3}Ej_iEj_k).
\end{equation}
For two arbitrary pure state components $\ket{\psi_i}$ and $\ket{\psi_k}$ in the optimal pure state
decomposition, we can obtain
\begin{eqnarray}\label{C2}
E1_i^2E1_k^2&\geq& (E2_i^2+E3_i^2)\cdot(E2_k^2+E3_k^2)\nonumber\\
&=& (E2_iE2_k)^2+(E3_iE3_k)^2+(E2_iE3_k)^2\nonumber\\
&&+(E3_iE2_k)^2\nonumber\\
&\geq& (E2_iE2_k)^2+(E3_iE3_k)^2+2(E2_iE2_k)\nonumber\\
&&\times(E3_iE3_k)\nonumber\\
&=&(\sum_{j=2}^3 Ej_iEj_k)^2,
\end{eqnarray}
where  we have used the monogamy relation for the pure-state case in the first inequality and the
perfect square trinomial equation in the second inequality. After taking the square root on both
sides of Eq. (\ref{C2}), we can get
\begin{equation}\label{C3}
E1_iE1_k-\sum_{j=2}^3 Ej_iEj_k\geq 0.
\end{equation}
Since $\ket{\psi_i}$ and $\ket{\psi_k}$ are two arbitrary components, the inequality in Eq. (\ref{C3})
is also satisfied for any other components, and thus the second term $\Delta\geq 0$ in Eq. (18).

In Eq. (20) of the main text, the second term has the same form as that in Eq. (18). Under the optimal
pure-state decomposition for $E_f(\rho_{A|B\mathbb{C}})$, we choose two arbitrary pure-state components
$\ket{\psi^i_{AB\mathbb{C}}}$ and $\ket{\psi^k_{AB\mathbb{C}}}$. After a  derivation similar to
those of Eqs. (\ref{C2}) and (\ref{C3}), we get $E1_iE1_k-\sum_{j=2}^3 Ej_iEj_k\geq 0$, where the pure-state
 SEF monogamy relation in $2\times 2\otimes 2^{N-2}$ systems is used. Because
$\ket{\psi^i_{AB\mathbb{C}}}$  and $\ket{\psi^k_{AB\mathbb{C}}}$ are two arbitrary components, we can
find that the second term in Eq. (20) is nonnegative.

\section{The calculation of bipartite multiqubit EOF in cavity-reservoir systems}

In Eq. (28) of the main text, the multipartite entanglement indicators in tripartite pure and mixed
states of cavity-reservoir systems have the forms
\begin{eqnarray}\label{D1}
\tau_{SEF(3)}(\ket{\Phi_t}_{c_1|c_2(r_1r_2)})&=&E_f^2(\ket{\Phi_t}_{c_1|c_2(r_1r_2)})
-E_f^2(\rho_{c_1c_2})\nonumber\\
&&-E_f^2(\rho_{c_1|r_1r_2}),\\
\tau_{SEF(3)}^{(2)}(\rho_{c_1|r_1r_2})&=&E_f^2(\rho_{c_1|r_1r_2})-E_f^2(\rho_{c_1r_1})\nonumber\\
&&-E_f^2(\rho_{c_1r_2}),
\end{eqnarray}
where the calculation of bipartite three-qubit EOF $E_f(\rho_{c_1|r_1r_2})$ is a key step for the application of
these indicators. According to the Koashi-Winter formula \cite{koashi04pra}, the EOF can be obtained via
quantum discord and we have
\begin{equation}\label{D3}
E_f(c_1|r_1r_2)=D(c_1|c_2)+S(c_1|c_2),
\end{equation}
where $S(c_1|c_2)=S(c_1c_2)-S(c_2)$ is the direct quantum generalization of conditional entropy
\cite{oll01prl}, with $S(\rho_x)=-\sum \lambda_i\mbox{log}_2\lambda_i$ being the von Neumann entropy,
and the quantum discord of two cavity photons is
\begin{equation}\label{D4}
D(c_1|c_2)=\mbox{min}_{\{E^{k}_{c_2}\}} \sum_k p_k S(c_1|E_{c_2}^k)-S(c_1|c_2),
\end{equation}
where the first term is the measurement-induced quantum conditional entropy \cite{oll01prl} with the
minimum runs over all the positive operator-valued measures. Chen \emph{et al.} presented an effective method for calculating
quantum discord and choosing optimal measurement \cite{che11pra}. After some analysis, we find that
the optimal measurement for the quantum discord $D(c_{1}|c_{2})$ is $\sigma_{x}$. Then, according to
Eq. (\ref{D3}), we can derive
\begin{equation}\label{D5}
E_{f}(\rho_{c_{1}|r_{1}r_{2}})=-\eta_{1}\mbox{log}_{2}\eta_{1}-(1-\eta_{1})
\mbox{log}_{2}(1-\eta_{1})
\end{equation}
where the parameter $\eta_{1}=[1-(1-4\beta^2 \xi^2 \chi^2)^{1/2}]/2$.

Similarly, for the multipartite entanglement indicators shown in Fig. 2(b)-2(d), we can
also derive the relevant multiqubit EOF via quantum discord. After some deduction, we can get
\begin{eqnarray}\label{D6}
E_{f}(\rho_{c_{1}|c_{2}r_{1}})=-\eta_{2}\mbox{log}_{2}\eta_{2}-(1-\eta_{2})
\mbox{log}_{2}(1-\eta_{2}),&&\nonumber\\
E_{f}(\rho_{r_{1}|c_{1}c_{2}})=-\eta_{3}\mbox{log}_{2}\eta_{3}-(1-\eta_{3})
\mbox{log}_{2}(1-\eta_{3}),&&\nonumber\\
E_{f}(\rho_{r_{1}|r_{2}c_{1}})=-\eta_{4}\mbox{log}_{2}\eta_{4}-(1-\eta_{4})
\mbox{log}_{2}(1-\eta_{4}),&&
\end{eqnarray}
with the parameters being
\begin{eqnarray}\label{D7}
\eta_{2}&=&\{1-[1-4\beta^2\xi^2(\beta^2+\xi^2-2\beta^2\xi^2)]^{1/2}\}/2,\nonumber\\
\eta_{3}&=&[1-(1-4\beta^2 \xi^2 \chi^2)^{1/2}]/2,\\
\eta_{4}&=&\{1-[1-4\beta^2\chi^2(\beta^2+\chi^2-2\beta^2\chi^2)]^{1/2}\}/2,\nonumber
\end{eqnarray}
respectively.

\section{Proof of theorem 4}

In a multipartite pure state of $2\otimes d_2\otimes d_3 \cdots \otimes d_{N-1}\otimes d_N$ systems,
the monogamy relation of the SEF is
\begin{eqnarray}\label{E1}
&&E_f^2(\ket{\Psi}_{A_1|\tilde{A}_2\cdots \tilde{A}_N})-\sum_{i} E_f^2(\rho_{A_1\tilde{A}_i})\nonumber\\
&=& E_f^2(C^2_{A_1|\tilde{A}_2\cdots \tilde{A}_N})-\sum_{i} E_f^2(C^2_{A_1\tilde{A}_i})\nonumber\\
&=& k_1 C^2_{A_1|\tilde{A}_2\cdots \tilde{A}_N}-\sum_i k_i C^2_{A_1\tilde{A}_i}\nonumber\\
&=& k_1(C^2_{A_1|\tilde{A}_2\cdots \tilde{A}_N}-\sum_i C^2_{A_1\tilde{A}_i})+\Gamma_1,
\end{eqnarray}
where the subscript $i\in\{2,N\}$ and we have used Theorem 1 in the main text in the first equality ;
 the relations $k_1\geq k_i$, with $k_1=E_f^2(C^2_{A_1|\tilde{A}_2\cdots
\tilde{A}_N})/C^2_{A_1|\tilde{A}_2\cdots \tilde{A}_N}$ and
$k_i=E_f^2(C^2_{A_1\tilde{A}_i})/C^2_{A_1\tilde{A}_i}$,in the second equality ; and the nonnegative
parameter $\Gamma_1=\sum_i (k_1-k_i) C^2_{A_1\tilde{A}_i}$ in the last equality.

When the SC is monogamous in the multipartite pure state $\ket{\Psi}_{A_1\tilde{A}_2\cdots
\tilde{A}_N}$, we have the parameter
\begin{equation}\label{E2}
\Gamma_2=k_1\cdot(C^2_{A_1|\tilde{A}_2\cdots \tilde{A}_N}-\sum_i C^2_{A_1\tilde{A}_i})\geq 0.
\end{equation}
Therefore, the monogamy relation in Eq. (\ref{E1}) is
\begin{eqnarray}\label{E3}
&&E_f^2(\ket{\Psi}_{A_1|\tilde{A}_2\cdots \tilde{A}_N})-\sum_{i} E_f^2(\rho_{A_1\tilde{A}_i})\nonumber\\
&&=\Gamma_1+\Gamma_2\geq 0,
\end{eqnarray}
since both parameters $\Gamma_i$ are nonnegative. Furthermore, for the mixed-state case,
we have
\begin{eqnarray}\label{E4}
&&E_f^2(\rho_{A_1|\tilde{A}_2\cdots \tilde{A}_N})-\sum_j E_f^2(\rho_{A_1\tilde{A}_j})\nonumber\\
&\geq& (\sum_i \mathcal{E}1_i)^2-\sum_j (\sum_i \mathcal{E}j_i)^2 \nonumber\\
&=& \sum_i (\mathcal{E}1_i^2-\sum_j \mathcal{E}j_i^2)+\Theta \geq 0,
\end{eqnarray}
where, in the first inequality, we have used the optimal pure-state decomposition
$\rho_{A_1\tilde{A}_2\cdots \tilde{A}_N}=\sum_i p_i \proj{\Psi^i}$ for $E_f(\rho_{A_1|\tilde{A}_2\cdots
\tilde{A}_N})$, with $\mathcal{E}1_i=p_i E_f(\ket{\Psi^i}_{A_1|\tilde{A}_2\cdots \tilde{A}_N})$, and
the relation $E_f(\rho_{A_1\tilde{A}_j})\leq \sum_i \mathcal{E}j_i$, with
$\mathcal{E}j_i=p_i E_f(\rho^i_{A_1\tilde{A}_j})$; in the second equality, the first term is nonnegative
due to the pure state monogamy property, and the second term
$\Theta=2\sum_i\sum_{k=i+1}(\mathcal{E}1_i\mathcal{E}1_k-\sum_{j=2}^{N}\mathcal{E}j_i\mathcal{E}j_k)$
is also nonnegative after an analysis similar to that in Appendix C. Therefore, we find that the
SEF is monogamous in the multipartite systems when the SC obeys this property.

Next, we consider the situation where the SC is polygamous in multipartite systems,
\begin{equation}\label{E5}
C^2(\rho_{A_1|\tilde{A}_2\cdots \tilde{A}_N})-\sum_j C^2(\rho_{A_1\tilde{A}_j})\leq 0,
\end{equation}
which results in the SC also being polygamous in the pure state case. In this case, we have
\begin{equation}\label{E6}
\Gamma_2=k_1\cdot(C^2_{A_1|\tilde{A}_2\cdots \tilde{A}_N}-\sum_i C^2_{A_1\tilde{A}_i})<0,
\end{equation}
and then the monogamy relation in Eq. (\ref{E1}) is
\begin{eqnarray}\label{E7}
E_f^2(\ket{\Psi}_{A_1|\tilde{A}_2\cdots \tilde{A}_N})-\sum_{i} E_f^2(\rho_{A_1\tilde{A}_i})
=\Gamma_1-|\Gamma_2|,
\end{eqnarray}
which is monogamous when the parameter $\Gamma_1$ is not less than the absolute value of the parameter $\Gamma_2$, \emph{i.e.},
\begin{equation}\label{E8}
\Gamma_1\geq |\Gamma_2|.
\end{equation}
Furthermore, when this monogamy relation of SEF in Eq. (\ref{E7}) is satisfied, we can find that
the mixed-state case holds via an analysis similar to that in Eq. (\ref{E4}). Thus, we find that the SEF
can be monogamous even if the SC is polygamous, and an example is shown in Eqs. (35) and (36) of
the main text.

Combining the cases with the SC being monogamous and polygamous, we find that in multipartite
$2\otimes d_2\otimes d_3 \cdots \otimes d_{N-1}\otimes d_N$ systems, the monogamy property of SEF
is superior to that of SC, which completes the proof of Theorem 4 in the main text.

\section{Monogamy properties of SEF and SC in a $4\otimes 2\otimes 2$ quantum state}

For the $4\otimes 2\otimes 2$ quantum state $\ket{\Phi}_{ABC}$ shown in Eq. (39) of the main text,
the bipartite reduced state for subsystem $AB$ can be written as
\begin{equation}\label{F1}
\rho_{AB}=\frac{1}{2}\proj{\varphi_1}+\frac{1}{2}\proj{\varphi_2},
\end{equation}
where the pure-state components are $\ket{\varphi_1}=\alpha\ket{00}+\beta\ket{11}$ and
$\ket{\varphi_2}=\alpha\ket{20}+\beta\ket{31}$, respectively. In an arbitrary pure state
decomposition of $\rho_{AB}$, the pure-state component has the form
\begin{equation}\label{F2}
\ket{\tilde{\varphi}_i}_{AB}=a_i\ket{\varphi_1}+e^{-i\gamma}\sqrt{1-a_i^2}\ket{\varphi_2}
\end{equation}
for which the reduced density matrix $\rho_B^i=\mbox{diag}\{\alpha^2,\beta^2\}$. Therefore, according to
the definition of EOF in Eq. (3) of the main text, we have
\begin{equation}\label{F3}
E_f(\rho_{AB})=S(B)=-\alpha^2\mbox{log}_2\alpha^2-\beta^2\mbox{log}_2\beta^2.
\end{equation}
Similarly, for the reduced quantum state $\rho_{AC}$, we have $E_f(\rho_{AC})=1$. Moreover, the reduced
quantum state of subsystem $A$ is $\rho_{A}=\mbox{diag}\{\alpha^2/2,\beta^2/2,\alpha^2/2,\beta^2/2\}$,
from which we get
\begin{equation}\label{F4}
E_f(\ket{\Phi}_{A|BC})=S(A)=S(B)+1.
\end{equation}
Thus, the monogamy property of SEF is
\begin{eqnarray}\label{F5}
M(SEF)&=&E_f^2(A|BC)-E_f^2(AB)-E_f^2(AC)\nonumber\\
&=&S^2(A)-S^2(B)-1^2\nonumber\\
&=&2S(B),
\end{eqnarray}
which is nonnegative, and therefore, the SEF is monogamous.

Next, we analyze the distribution of SC in this quantum state. For the bipartite $4\otimes 2$ mixed
state $\rho_{AB}$, its concurrence is defined by the convex roof extension \cite{ben96pra2,run01pra}
\begin{equation}\label{F6}
C(\rho_{AB})=\mbox{min}\sum_i p_i C(\ket{\psi^i}_{AB}),
\end{equation}
where the minimum runs over all the pure-state decompositions and the pure-state concurrence is
$C(\ket{\psi^i}_{AB})=\sqrt{2(1-\mbox{Tr}\rho_{B_i}^2)}$ \cite{kch05prl}. According to the
property of pure state decomposition in Eq. (\ref{F2}), we can derive
\begin{equation}\label{F7}
C^2_{AB}=4\alpha^2\beta^2.
\end{equation}
In a similar way, we can obtain $C^2_{AC}=1$. Moreover, the concurrence in the partition $A|BC$ is
\begin{equation}\label{F8}
C^2_{A|BC}=2(1-\mbox{Tr}\rho^2_A)=2-\alpha^4-\beta^4,
\end{equation}
and then the monogamy relation of SC is
\begin{eqnarray}\label{F9}
M(SC)&=&C^2_{A|BC}-C^2_{AB}-C^2_{AC}\nonumber\\
&=&(2-\alpha^4-\beta^4)-4\alpha^2\beta^2-1\nonumber\\
&=&-2\alpha^2\beta^2,
\end{eqnarray}
which is polygamous. In Fig. 4, the parameters are chosen to be $\alpha=\mbox{cos}\theta$ and
$\beta=\mbox{sin}\theta$, and the distributions $M(SEF)$ and $M(SC)$ are plotted as functions of
parameter $\theta$, which illustrates the different entanglement properties of SEF and SC.

\end{document}